\documentclass[preprint,showpacs,preprintnumbers,amsmath,amssymb]{revtex4}
\usepackage{epsfig}

\let\jnfont=\rm
\def\NPB#1,{{\jnfont Nucl.\ Phys.\ B }{\bf #1},}
\def\PLB#1,{{\jnfont Phys.\ Lett.\ B }{\bf #1},}
\def\EPJC#1,{{\jnfont Eur.\ Phys.\ Jour.\ C }{\bf #1},}
\def\PRD#1,{{\jnfont Phys.\ Rev.\ D }{\bf #1},}
\def\PRL#1,{{\jnfont Phys.\ Rev.\ Lett.\ }{\bf #1},}
\def\MPLA#1,{{\jnfont Mod.\ Phys.\ Lett.\ A }{\bf #1},}
\def\JPG#1,{{\jnfont J.\ Phys.\ G}{\bf #1},}
\def\CTP#1,{{\jnfont Commun.\ Theor.\ Phys.\ }{\bf #1},}
\def\ZPC#1,{{\jnfont Z.\ Phys.\ C }{\bf #1},}
\def\JHEP#1,{{\jnfont JHEP \ }{\bf #1},}

\begin{document}

\title{Lepton flavor violating Z-boson decays at GigaZ as a probe of supersymmetry}

\author{Jin Min Yang}
\affiliation{
     Institute of Theoretical Physics, Academia Sinica, Beijing 100190, China
     \vspace*{1.5cm}}

\begin{abstract}
We briefly review the lepton flavor violating Z-decays at GigaZ as
a probe of supersymmetry by focusing on $Z \to \ell_i
\overline{\ell}_j$ in two representative supersymmetric models:
the minimal supersymmetric model without $R$-parity and the
supersymmetric seesaw model. We conclude that under the current
experimental constraints from LEP and $\ell_i\to \ell_j \gamma$,
these rare decays can still be enhanced to reach the sensitivity
of the GigaZ. Therefore, supersymmetry can be probed via these
decays at GigaZ. \vspace*{1.5cm}

\end{abstract}


\maketitle

\section{INTRODUCTION}
The main task of particle physics in the current Large Hadron
Collider (LHC) era is probing new physics. The LHC is a powerful
discovery machine because of its high energy, but it is not an
ideal place for precision test of a theory because of its huge QCD
background. If new physics appears at TeV scale, as speculated and
expected by most theorists, the LHC will undoubtedly unveil it.
Then the proposed International Linear Collider (ILC) will take
the task of precision test of such new physics.

At the ILC the GigaZ option is expected to produce more than
$10^9$ Z-bosons \cite{ILC} and will play an important role in
probing new physics related to Z-boson. One sensitive probe is
through the flavor-changing neutral-current (FCNC) Z-boson decays
$Z \to \ell_i \overline{\ell}_j$, which are suppressed to be
unobservably small in the Standard Model (SM) but could be greatly
enhanced in new physics models like supersymmetry
\cite{cao-zll-rv,Z-decay,cao-zll-seesaw,zll-susy}.

In this review, we recapitulate the studies on the decays
$Z \to \ell_i \overline{\ell}_j$ in the $R$-parity violating minimal
supersymmetric model (RPV-MSSM) \cite{cao-zll-rv,Z-decay}
and the supersymmetric seesaw model \cite{cao-zll-seesaw}.
In Sec.\ref{sec2} we delineate the study in RPV-MSSM.
In Sec.\ref{sec3} we elucidate the study in the supersymmetric seesaw
model.
Finally, a summery is given in Sec. \ref{sec4}.

\section{Lepton flavor violating Z-decay in RPV-MSSM}
\label{sec2} In the MSSM the $R$-violating interactions are given
by
\begin{equation}\label{poten}
{\cal W}_{\not \! R}=\frac{1}{2}\lambda_{ijk}L_iL_jE_k^c
+\lambda'_{ijk} L_iQ_jD_k^c
+\frac{1}{2}\lambda''_{ijk}\epsilon^{abd}U_{ia}^cD_{jb}^cD_{kd}^c
+\mu_iL_iH_2,
\end{equation}
where $i,j,k$ are generation indices, $c$ denotes charge
conjugation, $a$, $b$ and $d$ are the color indices with
$\epsilon^{abd}$ being the total antisymmetric tensor,  $H_{2}$ is
the Higgs-doublet chiral superfield, and $L_i(Q_i)$ and
$E_i(U_i,D_i)$ are the left-handed lepton (quark) doublet and
right-handed lepton (quark) singlet chiral superfields. These
interactions have rich phenomenology which has been studied
intensively \cite{rp2} and a list of bounds is summarized in
\cite{review}.

The lepton flavor violating (LFV) processes, which are extremely
suppressed in the SM, may be greatly enhanced by these
$R$-violating interactions since both $\lambda$ and
$\lambda^{\prime}$ couplings can make contributions. Such
$R$-violating effects in the decays $Z \to \ell_i \bar\ell_j$ and
$\ell_i \to \ell_j \gamma$
 were studied in \cite{cao-zll-rv,Z-decay,liljgamma}.
Taking the presence of $\lambda_{ijk}'$ as an example, the LFV
interactions $\ell_i \bar\ell_j V$ ($V=\gamma,Z$) can be induced
at loop level by exchanging a squark $\tilde u^j_L$ or $\tilde
d^{k}_R$, as shown in Fig.1.
\begin{figure}[htbp]
\epsfig{file=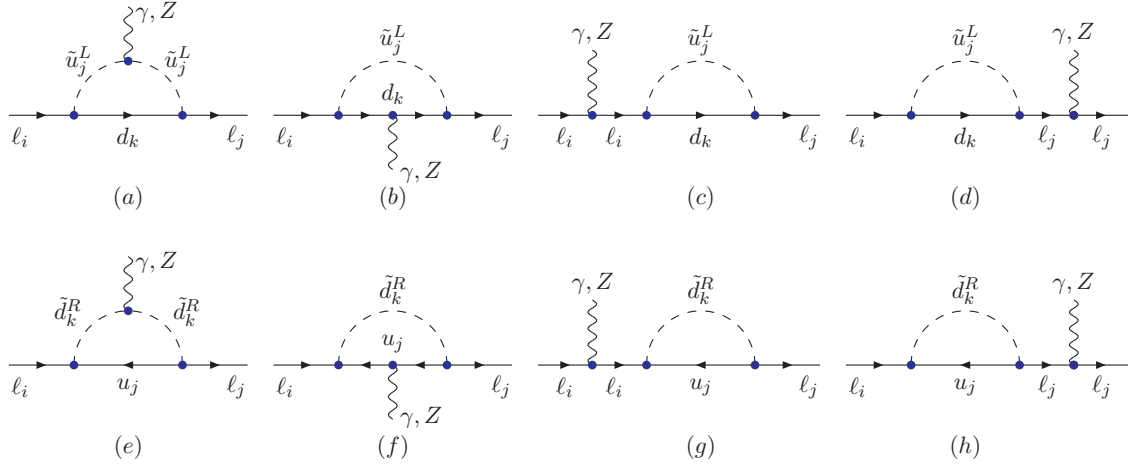,width=15cm} \vspace*{-0.7cm}
\caption{Feynman diagrams for $\ell_i - \ell_j $ transition induced by
        $L$-violating couplings at one-loop level.}
\label{fig1}
\end{figure}
So far the relevant constraints are from $\ell_i \to \ell_j \gamma$
given by \cite{exp4}
\begin{eqnarray}
BR(\mu \to e \gamma)&<&1.2\times 10^{-11},\\
BR(\tau \to e \gamma)&<&1.1\times 10^{-7},\\
BR(\tau \to \mu \gamma)&<& 4.5\times 10^{-8},
\label{bound}
\end{eqnarray}
and the LEP bounds on $Z \to \ell_i \bar\ell_j$ given by \cite{LEP1-LEP2}
\begin{eqnarray}
BR(Z \to \mu e)&<&1.7\times 10^{-6},\\
BR(Z \to \tau e)&<&9.8\times 10^{-6},\\
BR(Z \to \tau \mu)&<& 1.2\times 10^{-5}.
\end{eqnarray}
The possible sensitivity of GigaZ to the LFV decays of $Z$-boson
could reach \cite{gigaz}
\begin{eqnarray}
BR(Z \to \mu e)&\sim &2.0\times10^{-9},\\
BR(Z \to \tau e)&\sim &\kappa\times6.5\times10^{-8},\\
BR(Z \to \tau \mu)&\sim & \kappa\times2.2\times10^{-8}
\label{sensitivity}
\end{eqnarray}
with the factor $\kappa$ ranging from 0.2 to 1.0. In Fig. 2 we
take $\kappa=1.0$ to show the sensitivity of GigaZ in RPV-MSSM
compared with the bounds from $\ell_i \to \ell_j \gamma$ and the
$Z$-decays at LEP. We see that under the current experimental
constraints, the LFV $Z$-decays can still be enhanced to the
sensitivity of GigaZ. This implies that the GigaZ can further
strengthen the bounds on the relevant R-violating couplings in
case of un-observation.
\begin{figure}[tb]
\epsfig{file=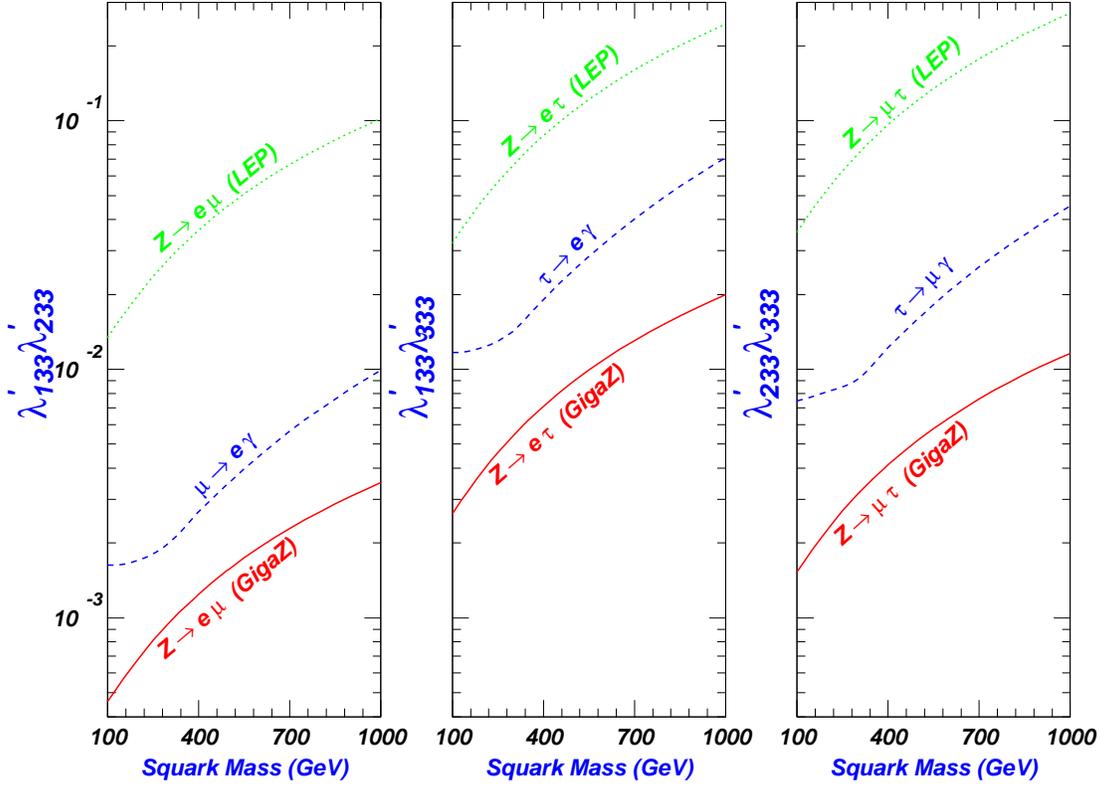,width=15cm} \vspace*{-0.7cm} \caption{The
$2\sigma$ sensitivity of lepton flavor vioalting $Z$-decays at
GigaZ in RPV-MSSM. Also shown are the bounds from $\ell_i \to
\ell_j \gamma$ and the $Z$-decays at LEP. These results are taken
from \cite{cao-zll-rv}.} \label{fig2}
\end{figure}

\section{Lepton flavor violating Z-decays in supersymmetric seesaw model}
\label{sec3} The seesaw mechanism \cite{Yanagida80} can be
realized in supersymmetric models by introducing right-handed
neutrino superfields with heavy Majorana masses
~\cite{Casas01-Hisano02}. In such a framework the flavor
diagonality of sleptons is usually assumed at the Planck scale,
but the flavor mixings at weak scale are inevitably generated
through renormalization equations since there is no symmetry to
protect the flavor diagonality. Such flavor mixings of  sleptons
generated at weak scale are proportional to neutrino Yukawa
coupling, which may be as large as top quark Yukawa coupling due
to see-saw mechanisim, and are enhanced by a large factor
$\log(M_P^2/M^2)$ ($M_P$ is Planck scale and $M$ is the neutrino
Majorana mass). Therefore, the popular mSUGRA with seesaw
mechanism predicts large flavor mixings of sleptons at weak scale.

With the right-handed neutrino superfields $\nu_R$,
the superpotential contains the $\nu_R$ terms
\begin{eqnarray}
W_\nu = -\frac{1}{2}\nu_R^{c}{\bf M}
\nu_R^c + \nu_R^{c} {\bf y_\nu} L \cdot H_2 \ , \label{eq1}
\end{eqnarray}
where ${\bf M}$ and ${\bf y_\nu}$ are matrices in flavor space,
$L$ and $H_2$ denote the left-handed lepton doublet and the Higgs
doublet with hypercharge $-1$ and $+1$, respectively.
The mass matrix of the charged sleptons is given by
\begin{eqnarray}
{\bf m}_{\tilde \ell}^2=\left(
    \begin{array}{cc}
        {\bf m}_{LL}^2    & {\bf m}_{LR}^{2\dag} \\
         {\bf m}_{LR}^2   & {\bf m}_{RR}^2
    \end{array}
      \right)
\label{slepmass}
\end{eqnarray}
with
\begin{eqnarray}
{\bf m}^2_{LL}&=&{\bf m}_{\tilde L}^2+\left[m_{\ell}^2+m_Z^2
\cos 2\beta \left(-\frac{1}{2}+s^2_W\right)\right]{\Large\bf 1} ,
\label{mlcharged} \\
{\bf m}^2_{RR}&=&{\bf m}_{\tilde R}^2+\left(m_{\ell}^2
- m_Z^2 \cos 2\beta\sin\theta_W^2\right){\Large\bf 1} ,\\
{\bf m}^{2}_{LR} &=&{\bf A}_\ell v\cos\beta-m_\ell\mu\tan\beta ~{\Large\bf 1} ,
\end{eqnarray}
where ${\Large\bf 1}$ is unit
$3\times 3$ matrix in generation space.

Since the right-handed sneutrinos have a mass as large as the
heavy Majorana neutrinos, their contributions to the LFV processes
can be ignored. Thus, only the left-handed sneutrinos are needed
to take into account, whose mass matrix is given by
\begin{eqnarray}
{\bf m}^2_{\tilde{\nu}}={\bf m}^2_{\tilde L}+\frac{1}{2}m_Z^2
       \cos 2\beta\ {\bf\Large 1} .
\end{eqnarray}
We assume universal soft-breaking parameters at the Planck scale:
\begin{eqnarray}
{\bf m}_{\tilde L}&=&{\bf m}_{\tilde R}=m_0\ {\bf\Large 1}, \\
\nonumber {\bf A}_\ell&=&A_0{\bf y}_\ell, \ \ {\bf A}_\nu=A_0{\bf
y}_\nu \ .
\end{eqnarray}
Since ${\bf y}_\ell$ and ${\bf y}_\nu$ cannot be diagonalized
simultaneously in general, it is usually assumed that ${\bf
y}_\ell$ is flavor diagonal but ${\bf y}_\nu$ is not. In this
basis the mass matrix of the charged sleptons is flavor diagonal
at Planck scale. However, when evolving down through
renormalization equations to weak scale, such flavor diagonality
is broken:
\begin{eqnarray} \label{eq:rnrges} \label{dmLij}
\delta ({\bf m}_{\tilde L}^2)_{IJ} &\simeq&
        -\frac{1}{8\pi^2}(3m_0^2+A_0^2)({\bf y}_\nu^{0\dag}{\bf y}_\nu^0)_{IJ}
        \ln\left(\frac{M_P}{\cal M}\right) \ , \\
\delta({\bf m}_{\tilde R}^2)_{IJ} &\simeq & 0 \ ,\\
  \label{smlreq}
\delta ({\bf A}_\ell)_{IJ} &\simeq& -\frac{3}{16\pi^2}A_0({\bf y}^0_\ell)_{II}
 ({\bf y}_\nu^{0\dag}{\bf y}_\nu^0)_{IJ} \ln\left(\frac{M_P}{\cal M}\right)\ ,
\end{eqnarray}
where ${\bf y}^0\equiv {\bf y}(M_P)$.
Therefore, both the charged sleptons and the left-handed sneutrinos
have mixings in flavor space. The flavor mixing of the charged sleptons
induces the FCNC couplings
$\tilde \chi^0_\alpha \ell_I \tilde\ell_J$ and $Z \tilde\ell_I
\tilde\ell_J$, while the flavor mixing of left-handed sneutrinos
induces the charged-current flavor-changing couplings
 $\tilde \chi^+_\alpha \ell_I \tilde\nu_J$.
These flavor-changing couplings
will contribute to the FCNC Z-decays  $Z\to\ell_i \bar\ell_j$,
as shown in Fig.3.
\begin{figure}[hbt]
\epsfig{file=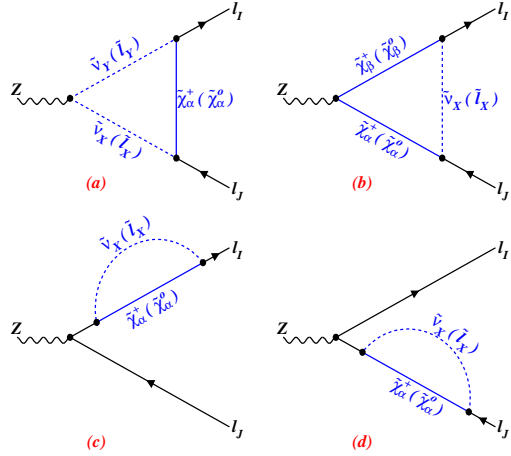,width=7cm} \vspace*{-0.5cm}
\caption{Feynman diagrams of SUSY contributions to the LFV processes
$Z\to\ell_i\bar{\ell}_j$. \label{Fig3}}
\end{figure}

With the constraints from current neutrino oscillation experiments
and introducing two right-handed neutrinos with masses ${\cal
M}_1=10^{13}$ GeV and ${\cal M}_2\simeq 10^{15}$ GeV, the
branching ratios of $Z\to \ell_i\bar{\ell}_j$ and
$\ell_i\to\ell_j\gamma$ versus the common scalar mass $m_0$ are
shown in Fig. 4. We see that the branching ratio of $Z\to \tau\mu$
can reach $10^{-8}$ in supersymmetric seesaw model (with the
current upper bound $BR(\tau \to \mu \gamma) < 4.5\times 10^{-8}$
shown in Eq.\ref{bound}, $Z\to \tau\mu$ with a branching ratio
$\sim 10^{-8}$ is allowed, as shown in  Fig. 4). Since the GigaZ
sensitivity for $Z\to \tau\mu$ is at $10^{-8}$, as shown in
Eq.(\ref{sensitivity}),  $Z\to \tau\mu$ may be accessible at GigaZ
and thus may serve as a probe of supersymmetric seesaw model.
\begin{figure}[htb]
\epsfig{file=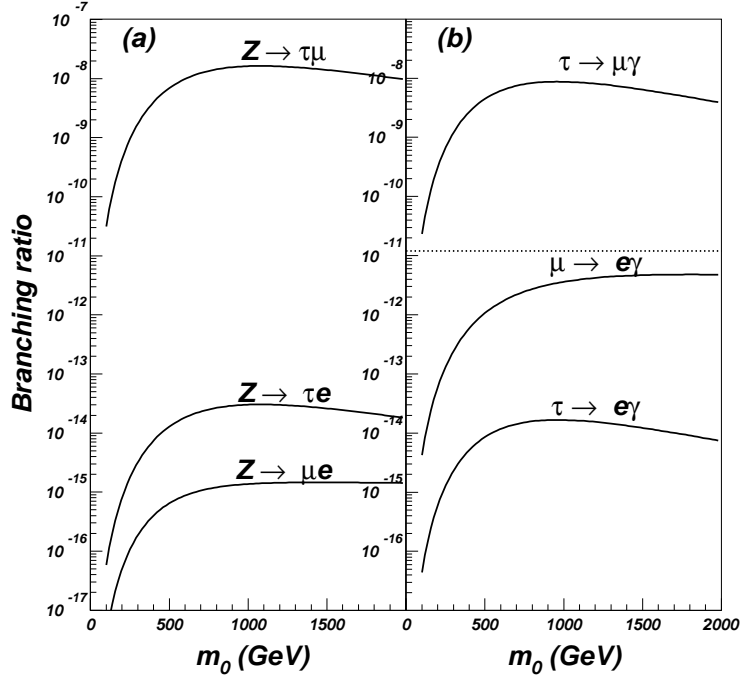,width=10cm}  \vspace*{-0.5cm}
\caption{Branching ratios of $Z\to \ell_i\bar{\ell}_j$ and
         $\ell_i\to\ell_j\gamma$ versus the common scalar mass
         $m_0$ \cite{cao-zll-seesaw}.}
\label{fig4}
\end{figure}

Note that while the above lepton flavor violating Z-decays serve
as a clean probe of new physics at the GigaZ, the FCNC decay modes
into quarks such as $Z \to b \bar s$ may also sensitive to new
physics.  In the SM  $Z \to b \bar s$ has a branching ratio of
$\sim 10^{-8}$ \cite{smzsb}, which could be greatly enhanced in
new physics models \cite{np-zsb}.

\section{CONCLUSION}
\label{sec4} From the lepton flavor violating Z-decays $Z \to
\ell_i \overline{\ell}_j$ in the $R$-parity violating minimal
supersymmetric model and the supersymmetric seesaw model, we
conclude that under the current experimental constraints from LEP
and $\ell_i\to \ell_j \gamma$, these decays can reach the
sensitivity of the GigaZ. Therefore, the supersymmetric models can
be probed via these decays at GigaZ.

\section*{Acknowledgement}
Most part of this review was finished while the author was a
visiting professor at Henan Normal University. This work was
supported by the National Natural Science Foundation of China
(NNSFC) under grant Nos. 10821504, 10725526 and 10635030.

\end{document}